\documentstyle[epsfig]{aipproc}

\begin{document}
\title{Nucleon-Nucleon interaction in a chiral constituent quark model\footnote{Talk presented by D. BARTZ at the International Workshop on Hadron Physics : ``Effective Theories of Low Energy QCD'', Coimbra, Portugal, September 10-15, 1999.}}

\author{D. Bartz and Fl. Stancu}
\address{University of Li\`ege, Institute of Physics, B.5, Sart Tilman, B-4000 Li\`ege 1, Belgium}

\maketitle

\begin{abstract}
We study the nucleon-nucleon (NN) problem as a six-quark system
in a nonrelativistic chiral constituent quark model where the
Hamiltonian contains a linear confinement and a pseudoscalar
meson (Goldstone boson) exchange interaction between the quarks.
This interaction has a long range Yukawa-type part, depending on the
mass of the exchanged meson and a short range part, mainly responsible
for the good description of the baryon spectra.
We calculate the NN potential in the adiabatic approximation as a
function of $Z$, the separation distance between the centres of
the two three-quark clusters. The orbital part of the six-quark
states is constructed either from the usual cluster model states or
from molecular orbital single particle states. The latter 
are more realistic, having proper axially and reflectionally 
symmetries. In both cases the potential presents an important
hard core at short distances, explained through the dominance 
of the $[51]_{FS}$ configuration. However in the molecular
orbital basis the core is less repulsive, as a 
consequence of the fact that this basis gives a better upper 
bound for the energy of the six-quark system.
We calculate the potential for the $^3S_1$ and $^3S_0$ channels
with two different parametrizations.  We find a small
(few MeV) attractive pocket for one of these parametrizations.
A middle range attraction is simulated by the addition of a
$\sigma$-meson exchange interaction between quarks, of a form
similar to that of the pseudoscalar meson exchange.
The present study is an intermediate, useful step towards
dynamical calculations based on the resonating group method.
\end{abstract}

\section*{Introduction}

There have been many attempts to study the nucleon-nucleon interaction starting
from a system of six interacting quarks described by a constituent quark model.
These models explain the short range repulsion as due to the colour magnetic
part of the one gluon exchange (OGE) interaction between quarks and due to
quark interchanges between two $3q$ clusters \cite{bartzOK84,bartzSH89}. To the OGE
interaction it was necessary to add a scalar and a pseudoscalar meson exchange
interaction
between quarks of different $3q$ clusters in order to explain the intermediate-
and long-range attraction between two nucleons \cite{bartzKU91,bartzZH94,bartzFU96}.\par
In \cite{bartzBA98} we have calculated the nucleon-nucleon ($NN$)
interaction potential at zero-separation distance between two three-quark
clusters in the frame of a constituent quark model \cite{bartzGL96a,bartzGL96b,bartzGL97a}
where the quarks interact via pseudoscalar meson i.e.
Goldstone boson exchange (GBE) instead of OGE. An
important motivation in using the GBE model is that it describes well the
baryon spectra. In particular, it correctly reproduces the order of positive
and negative parity states both for nonstrange \cite{bartzGL96b} and strange
\cite{bartzGL97a} baryons where the OGE model has failed.\par
The underlying symmetry of
the GBE model is related to the flavour-spin $SU_F(3) \times SU_S(2)$ group.
Combining it with the $S_3$ symmetry, a thorough analysis performed for the
$L=1$ baryons \cite{bartzCO98} has shown that the chiral quark picture leads 
to more satisfactory fits to the observed baryon spectrum than the OGE
models.\par
The one-pion exchange potential between quarks appears naturally as an
iteration of the instanton induced interaction in the t-channel \cite{bartzGV99}.
The meson exchange picture is also supported by explicit QCD
latice calculations \cite{bartzLIU98}.\par 
Another motivation in using the GBE model is that the exchange interaction
contains the basic ingredients required by the $NN$ problem. Its long-range
part, required to provide the long-range $NN$ interaction,
is a Yukawa-type potential depending on the mass of the exchange meson.
Its short-range part, of
opposite sign to the long-range one, is mainly responsible for the good
description of the baryon spectra \cite{bartzGL96a,bartzGL96b,bartzGL97a} and also induces a
short-range repulsion in the $NN$ system, both in the $^3S_1$ and $^1S_0$
channels \cite{bartzST97}. This study is an extention of \cite{bartzBA98} and 
we calculate here the interaction potential
between two $3q$ clusters as a function of $Z$, the separation distance between
the centres of the clusters. This separation distance is a good
approximation of the Jacobi relative coordinate between the two clusters. Under
this assumption, here we calculate the interaction potential in the adiabatic
(Born-Oppenheimer) approximation, as explained below.\par
A common issue in solving the $NN$ problem is the construction of adequate
six-quark basis states. The usual choice is a cluster model basis
\cite{bartzOK84,bartzSH89,bartzHA81}. In calculating the potential at zero-separation
distance,
in Ref. \cite{bartzBA98} we used molecular-type orbitals \cite{bartzST87} and
compared the
results with those based on cluster model single-particle states. The molecular
orbitals have the proper axially and reflectionally symmetries and can be
constructed from appropriate combinations of two-centre Gaussians. By using molecular orbitals, in Ref. \cite{bartzBA98} we found
that the height of the repulsion reduces by about 22 \% and 25 \% in the
$^3S_1$ and $^1S_0$ channels respectively with respect to cluster model
results.
It is therefore useful to analyse the role of molecular orbitals at distances
$Z \ne 0$. By construction, at $Z \rightarrow \infty$ the molecular orbital
states are simple parity conserving linear combinations of cluster model
states. Their role is expected to be important at short 
range at least. They also have the advantage of forming an orthogonal and
complete basis while the cluster model (two-centre) states are not orthogonal
and are overcomplete. For this reason we found that in practice they are more
convenient to be used than the cluster model basis, where one must carefully
\cite{bartzHA81} consider the limit $Z \rightarrow 0$.  Here too, for the purpose of
comparison we perform calculations both in the cluster and the molecular model.\par
In the following section we recall the procedure of constructing molecular orbital
single-particle states starting from the two-centre Gaussians used in the
cluster model calculations. Then the GBE Hamiltonian is presented. The 
subsequent section is devoted to results obtained for the $NN$ potential.
Next we introduce a middle range attraction through a scalar
meson exchange interaction between quarks parametrized consistently
with the pseudoscalar meson exchange. The last section is
devoted to a summary and conclusions.

\section*{Single-particle orbitals}
In the cluster model 
one can define states which in the limit of large
intercluster separation $Z$ are right $R$ and left $L$ states
\begin{equation}
R = \psi \left(\vec{r} - \frac{\vec{Z}}{2}\right) \hspace{8mm} \mbox{and}
\hspace{8mm} L =
\psi \left(\vec{r} + \frac{\vec{Z}}{2}\right).\label{bartzRLstates}
\end{equation}
In the simplest cluster model basis these are ground state harmonic oscillator
wave functions centered at $Z/2$ and $- Z/2$ respectively. They contain a
parameter $\beta$ which is fixed variationally to minimize the nucleon mass
described as a $3q$ cluster within a given Hamiltonian. The states (1) are
normalized but are not orthogonal at finite $Z$. They have good parity about
their centers but not about their common center $\vec{r} = 0$.\par
From $R$ and $L$ one constructs six-quark states of given orbital symmetry
$[f]_O$. The totally antisymmetric six-quark states also contain a flavour-spin
part of symmetry $[f]_{FS}$ and a colour part of symmetry $[222]_C$. In the
cluster model the most important basis states \cite{bartzST97} for the Hamiltonian
described in the following section are
\vspace{-0.1in}
\begin{equation}
\left|{\left.{{R}^{3}{L}^{3}\ {\left[{6}\right]}_{O}\
{\left[{33}\right]}_{FS}}\right\rangle}\right.\label{bartzClusSym1}
\end{equation}
\vspace{-0.35in}
\begin{equation}
\left|{\left.{{R}^{3}{L}^{3}\ {\left[{42}\right]}_{O}\
{\left[{33}\right]}_{FS}}\right\rangle}\right.\label{bartzClusSym2}
\end{equation}
\vspace{-0.35in}
\begin{equation}
\left|{\left.{{R}^{3}{L}^{3}\ {\left[{42}\right]}_{O}\
{\left[{51}\right]}_{FS}}\right\rangle}\right.\label{bartzClusSym3}
\end{equation}
\vspace{-0.35in}
\begin{equation}
\left|{\left.{{R}^{3}{L}^{3}\ {\left[{42}\right]}_{O}\
{\left[{411}\right]}_{FS}}\right\rangle}\right.\label{bartzClusSym4}
\end{equation}
Harvey \cite{bartzHA81} has shown that with a proper normalization the symmetry
$[6]_O$ contains only $s^6$ and $[42]_O$ only $s^4p^2$ configurations in the
limit $Z \rightarrow 0$.\par
According to Ref. \cite{bartzST87} let us consider also molecular orbital
single-particle states. 
Most generally these are eigenstates of a Hamiltonian $H_0$ having
axial and reflectional symmetries characteristic to the $NN$ problem. These
eigenstates have therefore good parity and good angular momentum projection. As
in the cluster model basis where one uses the two lowest states $R$ and $L$, in
the molecular orbital basis we also consider the two lowest states, $\sigma$,
of positive parity and $\pi$, of negative parity. From these we can construct
pseudo-right $r$ and pseudo-left $l$ states as
\begin{equation}
\left[ \begin{array}{c} r\\ l \end{array} \right] =
2^{-1/2}\, ( \sigma \pm \pi ) \hspace{5mm}
\rm{for\ all\ }Z,\label{bartzrlstates}
\end{equation}
where
\begin{eqnarray}
& <r|r> = <l|l> = 1 ,\,\, <r|l> = 0 .&\label{bartzrlorthonorm}
\end{eqnarray}
In principle one can obtain molecular orbital single particle states from mean
field calculations (see for example \cite{bartzKO94}). Here we approximate them by good
parity, orthonormal states constructed from the cluster model states (\ref{bartzRLstates}) as
\begin{equation}
\left[ \begin{array}{c} \sigma\\ \pi\end{array} \right] =
[2( 1 \pm <R|L>)]^{-1/2} ( R \pm L ) ,\label{bartzsigmapi}
\end{equation}
Such molecular orbitals are a very good approximation to the exact eigenstates
of a ``two-centre" oscillator frequently used in nuclear physics or
occasionally \cite{bartzRO87} in the calculation of the $NN$ potential. They provide
a convenient basis for the first step calculations based on the adiabatic
approximation as described below.\par
Introduced in (\ref{bartzrlstates}) they give
\begin{equation}
\left[ \begin{array}{c} r\\ l \end{array} \right] =
\frac{1}{2} \left[ \frac{R+L}{(1+<R|L>)^{1/2}} \pm
\frac{R-L}{(1-<R|L>)^{1/2}} \right].\label{bartzrlRLstates}
\end{equation}
At $Z \rightarrow 0$ one has $\sigma \rightarrow s$ and $\pi \rightarrow p$
(with $m = 0, \pm 1$) where $s$ and $p$ are harmonic oscillator states. Thus in
the limit $Z \rightarrow 0$ one has
\begin{equation}
\left[ \begin{array}{c} r\\ l \end{array} \right] =
2^{1/2} (s \pm p) ,\label{bartzrlspstates}
\end{equation}
and at $Z \rightarrow \infty$ one recovers the cluster model basis because $r
\rightarrow R$ and $\ell \rightarrow L$.\par
From $(r,l)$ as well as from $(\sigma,\pi)$ orbitals one can construct
six-quark states of required permutation symmetry, as shown in Ref. 
\cite{bartzST87}. In the limit $Z \rightarrow 0$ six-quark states obtained 
from molecular orbitals contain configurations of type $s^np^{6-n}$ with 
$n = 0,1,...,6$. For example the $[6]_O$ state contains $s^6$,
$s^6p^4$, $s^2p^4$ and $p^6$ configurations and the $[42]_O$ state
associated to
the $S$-channel contains $s^4p^2$ and $s^2p^4$ configurations. This is in
contrast to the cluster model basis where $[6]_O$ contains only $s^6$ and
$[42]_O$ only $s^4p^2$ configurations, as mentioned above.

Besides being poorer in $s^np^{6-n}$ configurations, the
number of basis states is smaller in the cluster model although we deal
with the
same $[f]_O$ and $[f]_{FS}$ symmetries and the same harmonic oscillator states
$s$ and $p$ in both cases. This is due to the existence of three-quark clusters
only
in the cluster model states, while the molecular basis also allows
configurations with five quarks to the left and one to the right, or vice
versa, or four quarks to the left and two to the right or vice versa (see
Eqs (11)-(19) of \cite{bartzBS99}). At large
separations these states act as ``hidden colour'' states but at 
short- and medium-range separation distances they are expected to
bring a significant contribution, as we shall see below.
The ''hidden colour" are states where a $3q$ cluster in an $s^3$ configuration
is a colour octet, in contrast to the nucleon which is a colour
singlet. Their role is important at short separations but it
vanishes at large ones
(see e.g. \cite{bartzHA81}).\par

\section*{Hamiltonian}
The GBE Hamiltonian considered in this study has the form \cite{bartzGL96b,bartzGL97a} :
\begin{equation}
H= \sum_i m_i + \sum_i \frac{\vec{p}_{i}^{\,2}}{2m_i} - \frac {(\sum_i
\vec{p}_{i})^2}{2\sum_i m_i} + \sum_{i<j} V_{\text{conf}}(r_{ij}) + \sum_{i<j}
V_\chi(r_{ij}) \, ,
\label{bartzham}
\end{equation}
with the linear confining interaction :
\begin{equation}
 V_{\text{conf}}(r_{ij}) = -\frac{3}{8}\lambda_{i}^{c}\cdot\lambda_{j}^{c} 
\, ( V_0 +C\, r_{ij} \,) ,
\label{bartzconf}
\end{equation}
and the spin--spin component of the GBE interaction in its $SU_F(3)$ form :
\begin{eqnarray}
V_\chi(r_{ij})
&=&
\left\{\sum_{F=1}^3 V_{\pi}(r_{ij}) \lambda_i^F \lambda_j^F \right.
\nonumber \\
&+& \left. \sum_{F=4}^7 V_{K}(r_{ij}) \lambda_i^F \lambda_j^F
+V_{\eta}(r_{ij}) \lambda_i^8 \lambda_j^8
+V_{\eta^{\prime}}(r_{ij}) \lambda_i^0 \lambda_j^0\right\}
\vec\sigma_i\cdot\vec\sigma_j,
\label{bartzVCHI}
\end{eqnarray}
\noindent
with $\lambda^0 = \sqrt{2/3}~{\bf 1}$, where $\bf 1$ is the $3\times3$ unit
matrix. The interaction (\ref{bartzVCHI}) contains $\gamma = \pi, K, \eta$ and $\eta '$
meson-exchange terms and the form of $V_{\gamma} \left(r_{ij}\right)$ is given
as the sum of two distinct contributions : a Yukawa-type potential containing
the mass of the exchanged meson and a short-range contribution of opposite
sign, the role of which is crucial in baryon spectroscopy.\par

In the parametrization of Ref. \cite{bartzGL96b} the exchange potential
due to a meson $\gamma$ has the form
\begin{equation}V_\gamma^{\rm{I}} (r)=
\frac{g_\gamma^2}{4\pi}\frac{1}{12m_i m_j}
\{\theta(r-r_0)\mu_\gamma^2 \,\frac{e^{-\mu_\gamma r}}{ r}- \frac {4}{\sqrt {\pi}}
\alpha^3 \exp(-\alpha^2(r-r_0)^2)\}.\label{bartzModel1}
\end{equation}

The shifted Gaussian of Eq. (\ref{bartzModel1}) results from a pure phenomenological fit
(see below) of the baryon spectrum with
\begin{equation}
r_0 = 0.43 \, { fm}, ~~\alpha = 2.91 \, { fm}^{-1},~~\label{bartzParamModel1a}
\end{equation}
\par
For the parametrization of Ref. \cite{bartzGL97a}, the potential has the form
\begin{equation}V_\gamma^{\rm{II}} (r)=
\frac{g_\gamma^2}{4\pi}\frac{1}{12m_i m_j}
\{\mu_\gamma^2 \,\frac{e^{-\mu_\gamma r}}{ r}-\lambda_\gamma^2 \,\frac{e^{-\lambda_\gamma r}}{ r}\}.\label{bartzModel2}
\end{equation}
where
\begin{equation}
\Lambda_\gamma = \Lambda_0+\kappa\ \mu_\gamma, ~~ 
\Lambda_0 = 5.82 \, { fm^{-1}}, ~~\kappa = 1.34.~~\label{bartzParamModel2a}
\end{equation}
\par
In the following, we shall call the form (\ref{bartzModel1}) Model I and the 
form (\ref{bartzModel2}) Model II. For a system of $u$ and $d$ quarks only, 
as it is the case here, the $K$-exchange does not contribute. The apriori 
determined parameters of the GBE model are the masses
\begin{equation}
m_{u,d} = 340 \, { MeV}, \,
\mu_{\pi} = 139 \, { MeV},~ \mu_{\eta} = 547 \, { MeV},~
\mu_{\eta'} = 958 \, { MeV}.\label{bartzParamModel1b}
\end{equation}
The other parameters are given in Table I.\par

\begin{table}\label{bartzI}\caption{Parameters of the Hamiltonian (\ref{bartzham}-\ref{bartzParamModel1b})}
\begin{tabular}{c|c|c|c|c|c}
\hline
Model & $V_0$ (MeV) & $C$ (fm$^{-2})$ & $g_8^2/4 \pi$ &
$g_0^2/4 \pi$ & Ref.\\
\hline
I & 0 & 0.474 & 0.67 & 1.206 & \cite{bartzGL96b}\\
\hline
II & -112 & 0.77 & 1.24 & 2.765 & \cite{bartzGL97a}\\
\end{tabular}
\end{table}


It is useful to  comment on Eqs. (\ref{bartzModel1}) and (\ref{bartzModel2}). 
The coupling of pseudoscalar mesons to 
quarks (or nucleons) gives rise to a two-body interaction potential which 
contains a Yukawa-type term and a contact term of opposite sign (see e.g. 
\cite{bartzBJ76}).  
The second term of (\ref{bartzModel1}) or (\ref{bartzModel2}) stems from the contact term, regularized with 
parameters fixed phenomenologically.  Certainly more fundamental studies 
are required to understand this second term and attempts are being made 
in this direction.  The instanton liquid model of the vacuum (for a review 
see \cite{bartzSS98}) implies point-like quark-quark 
interactions.  To obtain a realistic description of the hyperfine 
interaction this interaction has to be iterated in the t-channel \cite{bartzGV99}.  
The t-channel iteration admits a meson exchange interpretation \cite{bartzRB99}.
\section*{Results}
We diagonalize the Hamiltonian (\ref{bartzham})-(\ref{bartzParamModel1b}) in 
the six-quark cluster model basis and in the six-quark molecular orbital basis 
for values of the separation distance $Z$ up to 2.5 fm. 
Using in each case the lowest eigenvalue,
denoted by $\langle H\rangle_Z$ we define the $NN$ interaction potential in the
adiabatic (Born-Oppenheimer) approximation as
\begin{equation}
V_{NN}\left(Z\right) = \langle H\rangle_Z - 2m_N - K_{rel}\label{bartzadiab}
\end{equation}
Here $m_N$ is the nucleon mass obtained as a variational $s^3$ solution for a
$3q$ system described by the Hamiltonian (\ref{bartzham}). The wavefunction has the form
$\phi  \ \propto  \ \exp\ \ \left[{-\ \left({{\rho }^{2}\ +\ {\lambda
}^{2}}\right)/2{\beta }^{2}}\right]$ where $\rho =
\left(\vec{r}_1-\vec{r}_2\right)/\sqrt{2}$ and $\vec{\lambda} =
\left(\vec{r}_1+\vec{r}_2-2\vec{r}_3\right)/\sqrt{6}$. The minimum
for $m_N = \langle H\rangle_{3q}$ is 970 MeV in the Model I and 1311 MeV in the
Model II respectively, reached at the same $\beta$ in both models. The same 
value of $\beta$ is also used for the 6q system. This
is equivalent with imposing the ``stability condition'' which is of crucial
importance in resonating group method (RGM) calculations \cite{bartzOK84,bartzSH89}. The
quantity $K_{rel}$ represents the relative kinetic energy of two $3q$ clusters
separated at infinity
\begin{equation}
{K}_{rel}\ =\ {\frac{{3\hbar }^{2}}{4{m\beta }^{2}}}\label{bartzKrel}\ 
\end{equation}
where $m$ above and in the following designates the mass of the
$u$ or $d$ quark.
For our value of $\beta$ this gives $K_{rel} = 0.448$ GeV.
\begin{figure}[b!]
\centerline{\epsfig{file=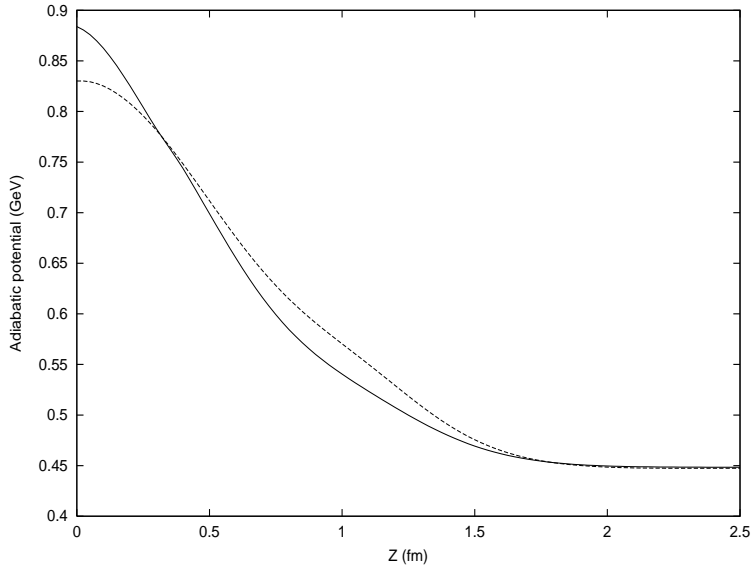,height=3in,width=4in}}
\vspace{10pt}
\caption{Comparison of the adiabatic potential in Model I, for $SI=(10)$, calculated in the cluster model (solid curve) and the molecular orbital basis (dashed curve).}
\end{figure}

\subsection*{Cluster model}

At $Z \rightarrow \infty$ the symmetries corresponding to
baryon-baryon channels, namely $[51]_{FS}$ and $[33]_{FS}$, must appear
with proper coefficients, as given by Eq. (\ref{bartzHarvey}). 
The contribution due to these symmetries
must be identical to the contribution
of $V_{\chi}$ to two nucleon masses also calculated with the Hamiltonian
(\ref{bartzham}). This is indeed the case. In the total Hamiltonian the contribution of the
$[411]_{FS}$ $V_{\chi}$ state tends to infinity when $Z \rightarrow \infty$. Then
this state decouples from the rest which is natural because
it does not correspond
to an asymptotic baryon-baryon channel. It plays a role at small $Z$
but at large $Z$ its amplitude in the $NN$ wavefunction vanishes, similarly to
the ``hidden colour" states. Actually, in diagonalizing the total Hamiltonian
in the basis (\ref{bartzClusSym1})-(\ref{bartzClusSym4}) we obtain an $NN$ wavefunction which in the limit $Z
\rightarrow \infty$ becomes \cite{bartzHA81}
\begin{equation}
{\psi }_{NN}\ =\ {\frac{1}{3}}\
\left|{\left.{{\left[{6}\right]}_{O}{\left[{33}\right]}_{FS}}\right\rangle}
\right.\
+\ {\frac{2}{3}}\
\left|{\left.{{\left[{42}\right]}_{O}{\left[{33}\right]}_{FS}}\right\rangle}
\right.\
-\ {\frac{2}{3}}\
\left|{\left.{{\left[{42}\right]}_{O}{\left[{51}\right]}_{FS}}\right\rangle}
\right.\label{bartzHarvey}
\end{equation}
The adiabatic potential drawn in Figs. 1 and
2 is defined according to Eq. (\ref{bartzadiab}) where
$\langle H\rangle_Z$ is the lowest eigenvalue resulting from the
diagonalization. Fig. 1 corresponds to $S$ = 1, $I$ = 0 in the Model I
and Fig. 2 to the same channel in the Model II. Note that from these curves
one should subtract $K_{rel}$ of Eq. (\ref{bartzKrel}) in order to obtain the 
asymptotic value zero for the potential.  Similar results are obtained in the 
$SI = (01)$ channel.
One can see that the potential is repulsive at any $Z$ in the Model I, but 
a small attractive pocket appears in the potential of the Model II (see the 
zoom in the inside box of Fig. 2).\par

\begin{figure}[t!]
\centerline{\epsfig{file=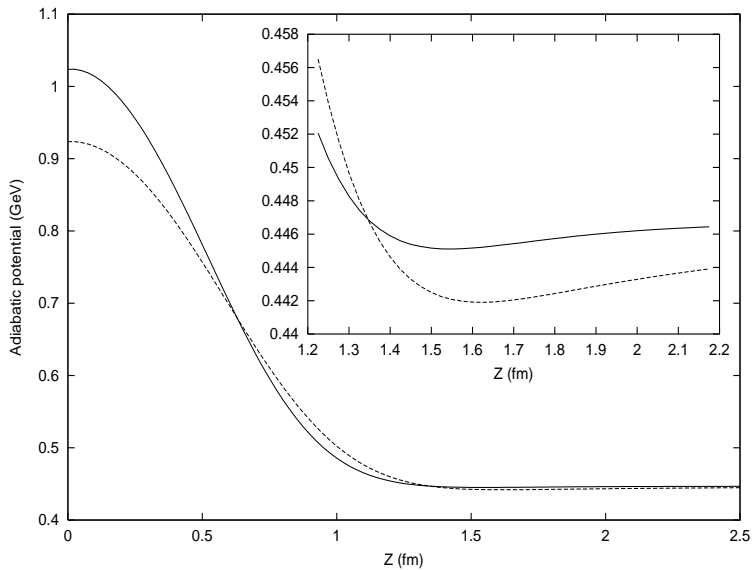,height=3in,width=4in}}
\vspace{10pt}
\caption{Same as Fig. 1 but for the Model II.  The upper box represents a zoom of the small attractive pocket.}
\end{figure}

\subsection*{Molecular orbital basis}

In the molecular
orbital basis the asymptotic form of  ${\psi }_{NN}$ is also given
by Eq. (\ref{bartzHarvey}) inasmuch as $r \rightarrow R$ and $l \rightarrow L$
as indicated below Eq. (\ref{bartzrlspstates}).\par

We diagonalize also the Hamiltonian and use its
lowest eigenvalue to obtain the $NN$ potential according to the definition
(\ref{bartzadiab}). The $S=1$, $I=0$ case is illustrated for Models I and II in
Figs. 1 and 2 
respectively, for a comparison with the cluster model basis. The repulsion 
reduces by about 15 \% in the $^3S_1$ channels when passing from the cluster
model basis to the molecular orbital basis. From Figs. 1 and 2 one can see 
that the molecular orbital basis has an important effect up to about 
$Z \approx$ 1.5 fm giving a lower potential at small values of $Z$.
For $Z \approx 1$ fm it gives a potential larger by few tens of MeV 
than the cluster model potential. However there is no attraction at all
in the Model I (for details see Ref. \cite{bartzBS99}).\par
\section*{Middle range attraction}
In principle we expected some attraction at large $Z$ due to the
presence of the Yukawa potential tail in Eq. (\ref{bartzModel1}). 
This is indeed the case in Model II. In Model I, we adopt a consistent 
procedure assuming that besides
the pseudoscalar meson exchange interaction there exists
an additional scalar, $\sigma$-meson exchange interaction between quarks.
This is in the spirit of the spontaneous chiral symmetry breaking 
mechanism on which the GBE model is based. The $\sigma$-meson is the 
chiral partner of the pion and it should be considered explicitly.\par

Actually once the one-pion exchange interaction between quarks is admitted,
one can inquire about the role of at least two-pion exchanges.
Recently it was found \cite{bartzRB99} that the two-pion exchange also
plays a significant role in the quark-quark interaction. It enhances the
effect of the isospin dependent spin-spin component of the one-pion
exchange interaction and cancels out its tensor component.
Apart from that it gives rise to a spin independent
central component, which averaged over the isospin wave function of
the nucleon it produces an attractive spin independent interaction.
These findings also support the introduction of a scalar ($\sigma$-meson)
exchange interaction between quarks as an approximate description
of the two-pion exchange loops.\par

For consistency with the parametrization \cite{bartzGL96b} we consider here
a scalar quark-quark interaction of the form
\begin{equation}V_\sigma (r)=
\frac{g_\sigma^2}{4\pi}\frac{1}{12m_i m_j}
\{\theta(r-r'_0)\mu_\sigma^2 \,\frac{e^{-\mu_\sigma r}}{ r}- \frac {4}{\sqrt {\pi}}
\alpha'^3 \exp(-\alpha'^2(r-r'_0)^2)\}.\label{bartzModelSigma}
\end{equation}
where $\mu_\sigma$ = 675 MeV and $r'_0$, $\alpha'$ and the coupling
constant $g^2_\sigma/4\pi$ are arbitrary parameters. In order
to be effective at medium-range separation between nucleons we expect
this interaction to have $r'_0 \neq r_0$ and $\alpha' \neq \alpha$.
Note that the factor $1/m_i m_j$ has only been introduced
for dimensional reasons.\par
\begin{figure}[b!]
\centerline{\epsfig{file=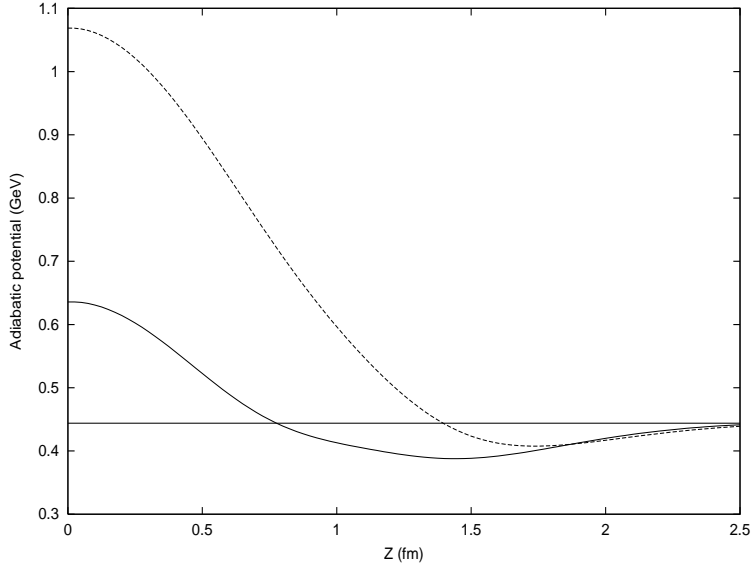,height=3in,width=4in}}
\vspace{10pt}
\caption{The adiabatic potential Model I in the molecular orbital basis for $SI=(10)$ (full curve) and $SI=(01)$ (dashed curve) with pseudoscalar + scalar quark-quark interaction.}
\end{figure}
We first looked at the baryon spectrum with the same variational 
parameters as before. 
The only 
modification is a shift of the whole spectrum which would correspond 
to taking $V_0 \approx - 60$ MeV in Eq. (\ref{bartzconf}).\par
For the $6q$ system we performed calculations in the molecular basis, 
which is more appropriate than the cluster model basis. We found that
the resulting adiabatic potential is practically insensitive to
changes in $\mu_\sigma$ and $r'_0$ but very sensitive to $\alpha'$.
In Fig. 3 we show results for 
\begin{equation}
r'_0 = 0.86 \, { fm}, ~\alpha'  = 1.47 \, { fm}^{-1},~~ g^2_\sigma/4\pi =
g^2_8/4\pi\label{bartzParamSigma}
\end{equation}
One can see that $V_\sigma$ produces indeed an attractive pocket,
deeper for $SI$ = (10) than for (01), as it should be for the
$NN$ problem. The depth
of the attraction depends essentially on $\alpha'$. The precise values
of the parameters entering Eq. (\ref{bartzModelSigma}) should be determined in 
further RGM calculations.  As mentioned above the 
Born-Oppenheimer potential is in fact the diagonal RGM kernel. 
It is interesting that an attractive pocket is seen in this kernel
when a $\sigma$-meson exchange interaction is combined 
with pseudoscalar meson exchange and OGE interactions (hybrid model),
the whole being fitted to the $NN$ problem. The same interaction should also 
be introduced in Model II, but here, our purpose is to show how the 
$\sigma$-exchange interaction gives some middle-range attraction.


\section*{Summary}
We have calculated the $NN$ potential in the adiabatic approximation 
as a function of $Z$, the separation distance between the centres of
the two $3q$ clusters. We used two different parametrizations of a 
constituent quark model where quarks interact via pseudoscalar
meson exchange. The orbital part of the six-quark states was constructed 
either from cluster model or molecular orbital single particle states.
The latter are more realistic, having the proper axially and reflectionally
symmetries. Also technically they are more convenient. We explicitly 
showed that they are important at small values of $Z$. In
particular we found that the
$NN$ potential obtained in the molecular orbital basis has a less repulsive
core than the one obtained in the cluster model basis. However 
none of the bases leads to an attractive pocket in one of the parametrizations
considered here. We have simulated
this attraction by introducing a $\sigma$-meson exchange interaction
between quarks.\par
The present calculations give us an idea about the size 
and shape of the hard core
produced by the GBE interaction.  Except for small values of $Z$ the 
two bases give rather similar potentials.  Taking $Z$ as a generator 
coordinate the following step is to perform
a dynamical study based on the resonating group method which
will provide phase-shifts to be compared to the experiment. These calculations
are underway.


\end{document}